\documentclass[amsmath,11pt]{article}

\usepackage{amssymb,amssymb,epsfig,bm}

\date{\empty}

\begin{document}

\title{\bf Relativistic approach to the kinematics of large-scale peculiar motions}
\author{Eleni Tsaprazi$^{1,2}$\footnote{\textit{Current address:} The Oscar Klein Centre, Department of Physics, Albanova University Center, Stockholm University, SE - 106 91 Stockholm, Sweden.}~~and Christos G. Tsagas$^{1,3}$\\ {\small $^1$Section of Astrophysics, Astronomy and Mechanics, Department of Physics}\\ {\small Aristotle University of Thessaloniki, Thessaloniki 54124, Greece}\\ {\small $^2$Sorbonne Universit\'es, UPMC Univ Paris 06, CNRS}\\ {\small Institut d'Astrophysique de Paris (IAP), 98bis Boulevard Arago, Paris 75014, France}\\ {\small $^3$DAMTP, Centre for Mathematical Sciences, University of Cambridge}\\ {\small Wilberforce Road, Cambridge CB3 0WA, UK}}

\maketitle

\begin{abstract}
We consider the linear kinematics of large-scale peculiar motions in a perturbed Friedmann universe. In so doing, we take the viewpoint of the ``real'' observers that move along with the peculiar flow, relative to the smooth Hubble expansion. Using relativistic cosmological perturbation theory, we study the linear evolution of the peculiar velocity field, as well as the expansion/contraction, the shear and the rotation of the bulk motion. Our solutions show growth rates considerably stronger than those of the earlier treatments, which were mostly Newtonian. On scales near and beyond the Hubble radius, namely at the long-wavelength limit, peculiar velocities are found to grow as $a^2$, in terms of the scale factor, instead of the Newtonian $a^{1/2}$-law. We attribute this to the fact that, in general relativity, the energy flux, triggered here by the peculiar motion of the matter, also contributes to the local gravitational field. In a sense, the bulk flow gravitates, an effect that has been bypassed in related relativistic studies. These stronger growth-rates imply faster peculiar velocities at horizon crossing and higher residual values for the peculiar-velocity field. Alternatively, one could say that our study favours bulk peculiar flows larger and faster than anticipated.
\end{abstract}

\section{Introduction}\label{sI}
Large-scale peculiar motions, also referred to as ``bulk flows'', are an established observational fact, confirmed by many surveys extending out to scales of several hundred Mpc (e.g.~see~\cite{Aetal} and references therein). Peculiar velocities are predicted by all structure formation scenarios, as the direct and unavoidable outcome of the ever increasing inhomogeneity and anisotropy of the post-recombination universe. The subject has a fairly long research history that goes back several decades. Nevertheless, although there are many structure-formation studies that incorporate peculiar velocities, essentially all the work that focuses on the evolution of the peculiar-velocity field is Newtonian, or quasi-Newtonian, in nature (see~\cite{P} and~\cite{M} respectively). In addition, to the best of our knowledge, all these studies are conducted in the rest-frame of the smooth Hubble expansion. The latter is defined as the coordinate system where the dipole in the Cosmic Microwave Background (CMB) spectrum vanishes. However, even at the linear level, there are subtle differences between the Newtonian and the relativistic treatments, given the very different way the two theories approach issues as fundamental as the time-vs-space relation and the nature of gravity itself. Moreover, no real observer in the universe follows the CMB frame, but we all move relative to it. Our Local Group of galaxies, for example, ``drifts'' with respect to the smooth Hubble flow at approximately 600~km/sec. For these reasons, the present work looks into the question of large-scale peculiar velocities and of their evolution by employing relativistic cosmological perturbation theory and by adopting the view point of observers living in typical galaxies (like our Milky Way) and moving relative to the smooth Hubble expansion. Put another way, all our calculations are done in a coordinate system moving with respect to the CMB frame. Our aim is to identify possible differences between the two approaches and thus provide a better theoretical understanding of the peculiar-motion kinematics, especially in view of the anticipated data from upcoming bulk-flow surveys. Note that the linear nature of our analysis implies that our scales of interest are large enough to neglect all nonlinear effects, which in practice means scales in excess of 100~Mpc.

We begin with a brief introduction to the analysis of peculiar velocities within the framework of relativistic cosmology in \S~\ref{sPVF}. In so doing, we assume a perturbed Friedmann-Robertson-Walker (FRW) universe filled with a pressureless fluid that can be baryonic, or low-energy cold dark matter (CDM), or both. We also adopt the 1+3~covariant and gauge invariant approach to cosmological perturbations (see~\cite{TCM,EMM} for recent reviews) and allow for two families of observers. These are the ``fictitious'' ideal observers that follow the CMB-frame and their ``real'' counterparts moving relative to it. The latter represent typical observers in our universe, living in galaxies like our Milky Way, which makes their coordinate system the natural frame to use in cosmological studies.

In addition to the peculiar velocity field itself, we investigate the full spectrum of the peculiar kinematics, namely the volume expansion/contraction, the shear and the rotation of the bulk motion. We establish the linear sources of cosmological peculiar motions in \S~\ref{sLSPFs}. As expected, we find that such flows are triggered by the growing inhomogeneity of the post-recombination universe. Our analysis also provides the full set of differential equations monitoring the peculiar velocity field, its expansion/contraction-rate its shear distortion and its rotation after decoupling. The homogeneous parts of these formulae accept analytic (power-law) solutions. These show increase, both in absolute terms and relative to the background Hubble expansion (see \S~\ref{sLEPFs}). More specifically, the peculiar velocity ($\tilde{v}$) is found to grow as $\tilde{v}\propto a^2$, where $a=a(t)$ is the cosmological scale-factor. This implies that the strength of the peculiar flow relative to the universal expansion, described by the ratio $\tilde{v}/v_H$ -- with $v_H$ representing the Hubble velocity on the corresponding length, grows as $\tilde{v}/v_H\propto a^{5/2}$. At the same time, the expansion/contarction-rate ($\tilde{\vartheta}$) of the bulk flow shows a scale-dependent growth, which reaches its maximum ($\tilde{\vartheta}\propto a$) on super-Hubble scales and tends to a constant (i.e.~$\tilde{\vartheta}\rightarrow$~constant) inside the horizon. Finally, the peculiar shear and the peculiar vorticity ($\tilde{\varsigma}$ and $\tilde{\varpi}$ respectively) are found to grow as $\tilde{\varsigma},\,\tilde{\varpi}\propto a$.

To the best of our knowledge, the growth rates given above are considerably stronger than those previously reported in the theoretical literature. The Newtonian/quasi-Newtonian treatments, in particular, lead to $v\propto a^{1/2}$ in the CMB frame~\cite{P,M}, which implies $v/v_H\propto a$ after decoupling. The same result was also recently obtained by applying the Newtonian version of the 1+3~formalism to the bulk-flow frame~\cite{TF}. The considerable difference between the relativistic and the Newtonian results stems from the fundamentally different way the two theories treat the gravitational field. More specifically, in relativity, the energy flux contributes to the stress-energy tensor of the matter and therefore to the local gravitational field (e.g.~see~\cite{TCM,EMM}). In our case, this purely relativistic effect ensures a flux-contribution to the stress-energy tensor due to the (peculiar) motion of the matter. Put another way, the bulk flow itself gravitates. This, in turn, feeds into the conservations laws and eventually appears in the equations monitoring cosmological perturbations and, more specifically, the evolution of peculiar-velocity perturbations (see \S~\ref{ssLRBTFs} and \S~\ref{ssPVs1} below). To our best knowledge, the relativistic effect reflecting the bulk-flow flux contribution to the local gravitational field, has been typically bypassed in related relativistic studies and the reader is referred to \S~\ref{ss4A} for further discussion and some characteristic examples.

Isolating and solving only the homogenous parts of the relativistic differential equations means that the resulting solutions are expected to hold on large enough scales, where the inhomogeneous components (consisting of spatial gradients) are negligible. With this in mind, the cautious approach should be to apply our power-law solutions on lengths near and beyond the Hubble radius, but not well inside the horizon. It may turn out that our long-wavelength results reduce to their Newtonian counterparts inside the Hubble scale, in which case the relativistic analysis could provide the ``initial conditions'' for the subsequent Newtonian treatments. In any case, stronger growth rates for the peculiar-velocity field on super-Hubble lengths, implies faster peculiar velocities at horizon-crossing, which in turn suggests higher residual values today. Therefore, at this stage, it is fair to say that the relativistic analysis seems to favour a number of surveys reporting bulk peculiar flows larger and faster than it is generally anticipated~\cite{WFH}.

\section{The peculiar velocity field}\label{sPVF}
In relativity neither time nor space retain their absolute Newtonian notion. Instead, observers moving with respect to each other have their own measure of time and space and they generally experience different versions of what we call ``reality''.

\subsection{The 4-velocity ``tilt''}\label{ss4-VT}
Let us consider two families of observers with (timelike) 4-velocities $u_a$ and $\tilde{u}_a$ respectively. Let us also assume that the latter family has peculiar velocity $\tilde{v}_a$ relative to the former. The three aforementioned velocity fields are then related by the familiar Lorentz boost
\begin{equation}
\tilde{u}_a= \tilde{\gamma}(u_a+ \tilde{v}_a)\,,  \label{4-vels}
\end{equation}
where $u_au^a=-1=\tilde{u}_a\tilde{u}^a$, $u_a\tilde{v}^a=0$, $\tilde{\gamma}=1/\sqrt{1-\tilde{v}^2}$ and $\tilde{v}^2= \tilde{v}_a\tilde{v}^a<1$ (see Fig.~\ref{fig:bflow}).\footnote{Throughout this manuscript Latin indices run from 0 to 3, while Greek ones take values from 1 to 3. Our metric tensor ($g_{ab}$) has signature $[-1,1,1,1]$ and we have also set $c=1$.} The ``tilt'' between the two 4-velocity vectors is determined by the (hyperbolic) angle $\beta$, defined by $\cosh\beta=-\tilde{u}_au^a= \tilde{\gamma}$. The latter ensures that $\beta=\ln(\tilde{\gamma}+ \sqrt{\tilde{\gamma}^2-1})$, with $\beta>0$ since $\tilde{\gamma}>1$ (see~\cite{KE} for further discussion and additional technical details).

\begin{figure}[tbp]
\centering \vspace{6cm} \includegraphics{bflow.eps} \caption{Observers ($O_1$, $O_2$) living inside the bulk-flow domain ($D$), moving with (local) peculiar velocity $\tilde{v}_a$ relative to the Hubble expansion.  The 4-velocities $u_a$ and $\tilde{u}_a$ define the CMB frame and the coordinate system of the peculiar motion respectively (see Eq.~(\ref{4-vels})).}  \label{fig:bflow}
\end{figure}

Note that, when dealing with large-scale bulk peculiar motions, the local and the mean velocities of the flow ($\tilde{v}$ and $\tilde{V}$ respectively) are typically related via the integral $\tilde{V}=(3/4\pi r^3)\int_{x<r}\tilde{v}{\rm d}x^3$, where $\tilde{v}^2=\tilde{v}_a\tilde{v}^a$ and $r$ is the (effective) radius of the bulk flow.

\subsection{The 1+3 threading}\label{ss1+3T}
Introducing two 4-velocities means that there are two temporal directions (along $u_a$ and $\tilde{u}_a$ respectively) and two 3-dimensional spatial sections (orthogonal to $u_a$ and $\tilde{u}_a$). Projecting onto these 3-spaces is achieved by using the symmetric projection tensors
\begin{equation}
h_{ab}= g_{ab}+ u_au_b \hspace{15mm} {\rm and} \hspace{15mm} \tilde{h}_{ab}= g_{ab}+ \tilde{u}_a\tilde{u}_b\,,  \label{3-spaces}
\end{equation}
with $h_{ab}u^a=0=\tilde{h}_{ab}\tilde{u}^b$ and $h_a{}^a=3=\tilde{h}_a{}^a$~\cite{TCM,EMM}. Note that, when there is no vorticity, these two projectors also act as the metric tensors of their associated 3-dimensional hypersurfaces.

We may now proceed to define temporal and spatial differentiation relative to the two timelike coordinate systems introduced in \S~\ref{ss4-VT} before. In particular, the time-derivatives in the $u_a$ and the $\tilde{u}_a$ frames are denoted by
\begin{equation}
^{\cdot}= u^a\nabla_a \hspace{15mm} {\rm and} \hspace{15mm} ^{\prime}= \tilde{u}^a\nabla_a\,,  \label{tmder}
\end{equation}
respectively. The corresponding spatial gradients, on the other hand, are
\begin{equation}
{\rm D}_a= h_a{}^b\nabla_b \hspace{15mm} {\rm and} \hspace{15mm} \tilde{\rm D}_a= \tilde{h}_a{}^b\nabla_b\,.  \label{spder}
\end{equation}
Using the above, one can decompose any spacetime variable, operator and equation into their timelike and spacelike components (relative to the $u_a$ or the $\tilde{u}_a$ field), thus achieving an 1+3 threading of the host spacetime into time and space~\cite{TCM,EMM}.

\subsection{Kinematic decomposition}\label{ssKD}
The irreducible kinematic variables of the two 4-velocity fields emerge by decomposing their gradients. More specifically, we have~\cite{Eh}
\begin{equation}
\nabla_bu_a= {1\over3}\,\Theta h_{ab}+ \sigma_{ab}+ \omega_{ab}- A_au_b\,.  \label{Nbua}
\end{equation}
In the above $\Theta={\rm D}^au_a$, $\sigma_{ab}={\rm D}_{\langle b}u_{a\rangle}$, $\omega_{ab}={\rm D}_{[b}u_{a]}$ and $A_a=\dot{u}_a$ are respectively the expansion/contraction scalar, the shear tensor, the vorticity tensor and the 4-acceleration vector of the $u_a$-field.\footnote{We use round brackets for symmetrisation, square for antisymmetrisation and angled brackets to denote symmetric and trace-free second-rank tensors. Consequently, $\sigma_{ab}={\rm D}_{\langle b}u_{a\rangle}={\rm D}_{(b}u_{a)}- ({\rm D}^cu_c/3)h_{ab}$ by default.} Positive values for $\Theta$ mean expansion, while in the opposite case we have contraction. Nonzero shear implies changes in the shape (under constant volume) of the associated fluid element, non-vanishing vorticity ensures rotation, while nonzero 4-acceleration indicates the presence of non-gravitational forces. In an exactly analogous way we may write
\begin{equation}
\nabla_b\tilde{u}_a= {1\over3}\,\tilde{\Theta}\tilde{h}_{ab}+ \tilde{\sigma}_{ab}+ \tilde{\omega}_{ab}- \tilde{A}_a\tilde{u}_b\,,  \label{tNbtua}
\end{equation}
for the gradient of the $\tilde{u}_a$-field. Note that, in cosmological studies, the volume scalar defines the scale factor of the universe by means of $\dot{a}/a=\Theta/3$. Also, the shear, the vorticity and the 4-acceleration are all spacelike, namely $\sigma_{ab}u^b=0=\omega_{ab}u^b=A_au^a$ by construction, with exactly analogous constraints applying to their ``tilded'' counterparts (i.e.~$\tilde{\sigma}_{ab}\tilde{u}^b=0= \tilde{\omega}_{ab}\tilde{u}^b=\tilde{A}_a\tilde{u}^a$).

No realistic fluid-flow is absolutely rigid, but instead it is expected to expand or contract, to change shape and to rotate, even by small amounts. It is therefore plausible to argue that large-scale bulk peculiar motions should behave in a similar manner. The expansion/contraction-rate, the shear distortion and the rotation of the $\tilde{v}_a$-field, technically speaking the irreducible variables of the peculiar kinematics, are obtained by splitting its spatial gradient. More specifically, taking the view point of the tilded observer, we have~\cite{ET}
\begin{equation}
\tilde{\rm D}_b\tilde{v}_a= {1\over3}\,\tilde{\vartheta}\tilde{h}_{ab}+ \tilde{\varsigma}_{ab}+ \tilde{\varpi}_{ab}\,,  \label{tDtv}
\end{equation}
with $\tilde{\vartheta}=\tilde{\rm D}^a\tilde{v}_a$, $\tilde{\varsigma}_{ab}=\tilde{\rm D}_{\langle b} \tilde{v}_{a\rangle}$ and $\tilde{\varpi}_{ab}=\tilde{\rm D}_{[b}\tilde{v}_{a]}$ representing the peculiar expansion/contraction, the peculiar shear and the peculiar vorticity respectively.\footnote{The scalar $\tilde{\vartheta}$ monitors the volume expansion/contraction of the bulk motion. When $\tilde{\vartheta}>0$ the peculiar flow expands and in the opposite case it contracts. The peculiar shear follows changes in the shape of the bulk flow, say from spherical to ellipsoidal, while the vorticity contains information about its rotation. At the linear level $|\tilde{\vartheta}|/\Theta\ll1$, $\tilde{\varsigma}/\Theta\ll1$ and $\tilde{\varpi}/\Theta\ll1$, with $2\tilde{\varsigma}^2= \tilde{\varsigma}_{ab}\tilde{\varsigma}^{ab}$ and $2\tilde{\varpi}^2=\tilde{\varpi}_{ab}\tilde{\varpi}^{ab}$.}

\subsection{Linear relations between the two frames}\label{ssLRBTFs}
So far we have not imposed any constraints on the peculiar velocity, which means that our definitions and our formulae hold for arbitrarily fast relative motions in a general spacetime. Hereafter, we will only consider non-relativistic drift velocities with $\tilde{v}^2\ll1$ and
\begin{equation}
\tilde{u}_a\simeq u_a+ \tilde{v}_a\,,  \label{nr4-vels}
\end{equation}
since $\tilde{\gamma}\simeq1$.\footnote{Typical Newtonian studies, define peculiar velocities by introducing physical and comoving coordinates ($r^{\alpha}$ and $x^{\alpha}$ respectively, with $\alpha=1,2,3$), related by $r^{\alpha}=ax^{\alpha}$. The time derivative of the latter leads to $v^{\alpha}=v_H^{\alpha}+ v_p^{\alpha}$, where $v^{\alpha}=\dot{r}^{\alpha}$ is the physical (total) velocity, $v_H^{\alpha}=Hr^{\alpha}$ is the Hubble velocity and $v_p^{\alpha}=a\dot{x}^{\alpha}$ is the peculiar velocity. The above given relation between the three velocity fields is the Newtonian analogue of Eqs.~(\ref{4-vels}) and (\ref{nr4-vels}). Having said that, we remind the reader that the 4-velocities $\tilde{u}_a$ and $u_a$ are both timelike vectors, whereas their Newtonian counterparts are all purely spatial.} Treating the peculiar velocity field as a perturbation, we assume that the host spacetime is an almost-FRW universe. Finally, we identify the $u_a$-field with the coordinate system of the Hubble flow and place the tilded observers in a typical galaxy like our Milky Way.\footnote{Although relativity postulates the absence of preferred coordinate systems, the universal expansion naturally selects the CMB frame as the reference system relative to which peculiar velocities should be defined and measured.} Then, the kinematic variables defined in \S~\ref{ssKD} are related by the linear expressions~\cite{M}
\begin{eqnarray}
\tilde{\Theta}= \Theta+ \tilde{\vartheta}\,, \hspace{15mm} \tilde{\sigma}_{ab}= \sigma_{ab}+ \tilde{\varsigma}_{ab}\,, \hspace{15mm} \tilde{\omega}_{ab}= \omega_{ab}+ \tilde{\varpi}_{ab} \label{linrels1a}
\end{eqnarray}
and
\begin{equation}
\tilde{A}_a= A_a+ \tilde{v}_a^{\prime}+ {1\over3}\,\Theta\tilde{v}_a\,,  \label{linrels1b}
\end{equation}
where $\tilde{v}_a^{\prime}=\tilde{u}^b\nabla_b\tilde{v}_a$ is the time derivative of the peculiar velocity in the tilded frame (not to be confused with the ``peculiar gravitational acceleration'' of the Newtonian treatments). Of particular interest for our purposes is Eq.~(\ref{linrels1b}). According to this relation, we cannot set both 4-acceleration vectors to zero simultaneously. When $A_a$ vanishes in the CMB frame, for example, the tilded observers measure a nonzero 4-acceleration solely because of their peculiar motion (i.e.~$\tilde{A}_a=\tilde{v}^{\prime}_a+(\Theta/3)\tilde{v}_a$ when $A_a=0$).

In a similar manner, one can show that the dynamical variables measured in the two frames are related by
\begin{equation}
\tilde{\rho}= \rho\,, \hspace{15mm} \tilde{p}= p, \hspace{15mm} \tilde{q}_a= q_a- (\rho+p)\tilde{v}_a \label{linrels2a}
\end{equation}
and
\begin{equation}
\tilde{\pi}_{ab}= \pi_{ab}\,,  \label{linrels2b}
\end{equation}
to first approximation~\cite{M}. Here, $\rho$ is the energy density, $p$ is the isotropic pressure, $q_a$ is the energy flux and $\pi_{ab}$ is the viscosity of the matter as measured in the $u_a$-frame, while their tilded counterparts are associated with the $\tilde{u}_a$-field. Expression (\ref{linrels2a}c) has special significance in this study, since it ensures an additional energy-flux contribution as a result of relative motion alone. In general relativity, the energy flux gravitates as well, since it contributes to the stress-energy tensor of the matter. Therefore, the peculiar motion of the matter also contributes to the local gravitational field. In a sense, the bulk flow itself gravitates.

In what follows we will use the above linear relations to study the kinematic evolution of large-scale peculiar motions in a perturbed Friedmann universe filled with pressureless matter (baryonic or/and CDM).

\section{Linear sources of peculiar flows}\label{sLSPFs}
Large-scale peculiar motions are treated as a result of the increasing inhomogeneity and anisotropy of our universe, due to the ongoing structure-formation process. The latter starts in earnest after recombination, once the baryons have decoupled from the background radiation field.

\subsection{Peculiar velocities}\label{ssPVs1}
Let us assume an almost-FRW cosmology filled with pressureless matter. This can be baryonic, or low-energy CDM, or a mixture of both. The assumption of an unperturbed FRW background, ensures that all perturbations (including those in the peculiar-velocity field) vanish there. This makes our analysis gauge invariant and therefore free of any gauge-related issues, in line with the Stewart \& Walker lemma~\cite{SW}. Also, the absence of pressure means that we can set $A_a=0$ in the CMB frame.\footnote{Setting $A_a=0$ means that the worldlines of the CMB frame are timelike geodesics, whereas those of the tilted observers are not. In principle, one could also set $\tilde{A}_a=0$, in which case $A\neq0$ (e.g.~see~\cite{M}). Then, it is the tilted observers that move along timelike geodesics, while their Hubble-flow counterparts follow non-geodesic lines. Here, we are assuming that the universe is an FRW model relative to the CMB frame, which sets $A_a=0$ by default. That aside, one can perform the analysis in either coordinate system and reach the same results.} Then, relation (\ref{linrels1b}), which also serves as the linear propagation equation of the peculiar velocity in the bulk-flow frame, recasts as
\begin{equation}
\tilde{v}_a^{\prime}= -H\tilde{v}_a+ \tilde{A}_a\,,  \label{tv'1}
\end{equation}
with $H=\Theta/3$ being the Hubble parameter in the CMB frame. Therefore, at the linear level, the sole source of peculiar velocities is the 4-acceleration. This makes $\tilde{A}_a$ the key to the subsequent evolution of the $\tilde{v}_a$-field. It also means that the way the 4-acceleration is treated is crucial. In the Newtonian analogue of Eq.~(\ref{tv'1}), the role of the 4-acceleration is played by the gradient of the gravitational potential (e.g.~see~\cite{P,TF}). Similarly, the quasi-Newtonian approach (which was the first -- to the best of our knowledge -- 1+3 covariant study of the issue) introduces an effective gravitational potential to account for the effects of the 4-acceleration~\cite{M}. The latter, however, pre-assumes that the perturbed spacetime is both irrotational and shear-free. Moreover, the necessary propagation formula of the 4-acceleration was obtained after introducing an ansatz for the time evolution of the aforementioned potential~\cite{M}. Here, we will take an alternative route. More specifically, by applying relativistic linear cosmological perturbation theory to the tilded frame, we will obtain analytical expressions for both the 4-acceleration vector and its time-derivative.

\subsection{The 4-acceleration}\label{ss4A}
Our starting point is the linear transformation law (\ref{linrels2a}c), which guarantees that, even when the cosmic medium appears as a perfect fluid in the CMB frame, there is an energy-flux vector in the tilded frame solely due to the latter's relative motion. More specifically, following (\ref{linrels2a}c), we deduce that $\tilde{q}_a=-\rho\tilde{v}_a$ when $q_a=0$. Consequently, there is a flux-contribution to the stress-energy tensor triggered by the peculiar motion of the matter, which then feeds into the energy and the momentum conservation laws and eventually reaches the evolution formulae of cosmological perturbations. In particular, when linearised in the tilded frame, the evolution formula of density inhomogeneities (e.g.~see Eqs.~(2.3.1) and (10.101) in~\cite{TCM,EMM}), reads~\cite{TK}
\begin{equation}
\tilde{\Delta}_a^{\prime}= -\tilde{\mathcal{Z}}_a+ {3aH\over\rho}\left(\tilde{q}_a^{\prime}+4H\tilde{q}_a\right)- {a\over\rho}\,\tilde{\rm D}_a\tilde{\rm D}^b\tilde{q}_b\,.  \label{tDelta'}
\end{equation}
Here, $\tilde{\Delta}_a=(a/\rho)\tilde{\rm D}_a\rho$ and $\tilde{\mathcal{Z}}_a=a\tilde{\rm D}_a\tilde{\Theta}$ to first approximation, representing inhomogeneities in the matter distribution and in the universal expansion respectively~\cite{TCM,EMM}. Employing relations (\ref{linrels1b}) and (\ref{linrels2a}c), with $A_a=0=q_a$ and $p=0$, it is straightforward to show that
\begin{equation}
\tilde{A}_a= -{1\over\rho} \left(\tilde{q}^{\prime}_a+4H\tilde{q}_a\right)\,.  \label{tAa1}
\end{equation}
Keeping in mind that $\tilde{q}_a=-\rho\tilde{v}_a$, the latter combines with Eq.~(\ref{tDelta'}) to give
\begin{equation}
\tilde{A}_a= {1\over3H}\,\tilde{\rm D}_a\tilde{\vartheta}- {1\over3aH}\left(\tilde{\Delta}^{\prime}_a +\tilde{\mathcal{Z}}_a\right)\,.  \label{tAa2}
\end{equation}
Note that in a Newtonian perturbative study Eq.~(\ref{tDelta'}) reduces to $\tilde{\Delta}^{\prime}_{\alpha}=- \tilde{\mathcal{Z}}_{\alpha}$, with $\tilde{\Delta}_{\alpha}= (a/\rho)\partial_{\alpha}\rho$ and $\tilde{\mathcal{Z}}_{\alpha}= a\partial_{\alpha}\tilde{\Theta}$~\cite{E}. The absence of an acceleration term in the above, explains why expression (\ref{tAa2}) has no close Newtonian analogue. There, as well as in the quasi-Newtonian treatments, the acceleration and the 4-acceleration are given by the gradient of the gravitational potential. All these make (\ref{tAa2}) the key ``relativistic correction'' and the practical reason for the differences between the relativistic and the Newtonian/quasi-Newtonian results.

The literature also contains a number of relativistic structure-formation studies. To the best of our knowledge, these investigations have not encountered the role of the bulk-flow flux in the evolution of peculiar-velocity perturbations, as reported here. The reasons vary and the following examples are indicative of that. Technically speaking, closer to our analysis are perhaps the multi-fluid scenarios, where the species involved have their individual peculiar velocities. These studies, however, are typically performed in the energy frame (otherwise known as Landau-Lifshitz frame), where the total flux vector is set to zero (i.e.~the individual fluxes cancel each other out -- e.g.~see~\cite{TCM,EMM} and also~\cite{CMV-R}). Then, there is no flux contribution to the stress-energy tensor and the relativistic effects of the bulk-flow motion described here are bypassed.\footnote{Ours is a single-fluid study. Nevertheless, one can still recover the results of the multi-fluid analysis ``phenomenologically'', by setting $\tilde{q}_a=0$ in Eq.~(\ref{tAa1}) and therefore $\tilde{A}_a=0$ in (\ref{tv'1}). The latter then leads to $\tilde{v}_a\propto a^{-1}$ and to $\tilde{\vartheta}\propto a^{-2}$, as in~\cite{CMV-R} and~\cite{TCM,EMM} respectively. Taken at face value, these solutions imply that the peculiar flows decay quickly with the universal expansion on all scales, which does not seem to agree with the observations.} There are also non-covariant relativistic approaches in the literature. These are usually gauge-dependent, with some of them adopting the comoving gauge, where the peculiar velocities vanish by default (e.g.~see~\cite{RV}). Other treatments allow for peculiar-velocity perturbations, but their evolution is not the focal point of the study. As a result, no explicit solutions are provided (e.g.~see~\cite{TK,MBBM}).\footnote{Assuming the direct comparison between Eq.~(\ref{tv'1}) here and relation (5.6) in~\cite{MBBM} is allowed, our 4-acceleration vector can be expressed as a sum of zero and first-order terms in the post-Friedmann approximation scheme.} Overall, there are relatively few analytical treatments focusing on the kinematics of cosmological peculiar motions. Moreover, essentially all the available studies are performed in the CMB and not in the tilded frame, which is by construction the natural coordinate system to analyse peculiar flows.

Before closing this section, we remind the reader that the 4-acceleration is the only source of peculiar-velocity perturbations (see Eq.~(\ref{tv'1}) in \S~\ref{ssPVs1}). Given that $\tilde{\vartheta}$ vanishes in the absence of drift motions, expression (\ref{tAa2}) leaves $\tilde{\Delta}_a^{\prime}$ and $\tilde{\mathcal{Z}}_a$ as the sole sources of linear peculiar velocities. Therefore, as expected, the $\tilde{v}_a$-field is induced by the increasing inhomogeneity of the post-recombination universe and more specifically by temporal variations in the density-gradients and by spatial variations in the universal expansion (described by $\tilde{\Delta}_a^{\prime}$ and $\tilde{\mathcal{Z}}_a$ respectively).

\section{Linear evolution of peculiar flows}\label{sLEPFs}
Expression (\ref{tv'1}) requires an evolution formula for the 4-acceleration to close the system of the differential equations. In what follows, we will do so and also obtain analytic solutions for the $\tilde{v}_a$-field.

\subsection{The key differential equations}\label{ssKDEs}
Taking the time derivative of Eq.~(\ref{tAa2}), recalling that $\dot{H}=-H^2[1+(\Omega/2)]$ in the FRW background (with $\Omega=\kappa\rho/3H^2$ being the associated density parameter) and using the linear commutation law $(\tilde{\rm D}_a \tilde{\vartheta})^{\prime}=\tilde{\rm D}_a \tilde{\vartheta}^{\prime}-H\tilde{\rm D}_a\tilde{\vartheta}$ between the temporal and the spatial derivatives of first-order scalars~\cite{M}, provides the propagation formula
\begin{equation}
\tilde{A}_a^{\prime}= {1\over2}\,H\Omega\left(\tilde{v}_a^{\prime} +H\tilde{v}_a\right)+ {1\over3H}\,\tilde{\rm D}_a \tilde{\vartheta}^{\prime}- {1\over3aH} \left(\tilde{\Delta}_a^{\prime\prime} +\tilde{\mathcal{Z}}_a^{\prime}\right)\,,  \label{tA'2}
\end{equation}
which monitors the linear evolution of the 4-acceleration relative to the bulk-flow frame. We therefore have analytic linear expressions for both the 4-acceleration and its time derivative, while avoiding the restrictions of the quasi-Newtonian analysis (see \S~\ref{ssPVs1} previously).

Substituting the right-hand side of (\ref{tA'2}) into the time-derivative of Eq.~(\ref{tv'1}) leads to
\begin{equation}
\tilde{v}_a^{\prime\prime}= -H\left(1-{1\over2}\,\Omega\right)\tilde{v}_a^{\prime}+ H^2\left(1+\Omega\right)\tilde{v}_a+ {1\over3H}\,\tilde{\rm D}_a\tilde{\vartheta}^{\prime}- {1\over3aH}\left(\tilde{\Delta}_a^{\prime\prime}+ \tilde{\mathcal{Z}}_a^{\prime}\right)\,.  \label{tv''2}
\end{equation}
In addition, using the linear commutation laws $\tilde{\rm D}_b \tilde{v}_a^{\prime}=(\tilde{\rm D}_b\tilde{v}_a)^{\prime}+ H\tilde{\rm D}_b\tilde{v}_a$ and $\tilde{\rm D}_v \tilde{v}_a^{\prime\prime}=(\tilde{\rm D}_b \tilde{v}_a)^{\prime\prime}+2H(\tilde{\rm D}_b\tilde{v}_a)^{\prime}- (H^2\Omega/2)\tilde{\rm D}_b\tilde{v}_a$, the linearised 3-gradient of the above reads
\begin{eqnarray}
\left(\tilde{\rm D}_b\tilde{v}_a\right)^{\prime\prime}&=& -3H\left(1-{1\over6}\,\Omega\right)\left(\tilde{\rm D}_b \tilde{v}_a\right)^{\prime}+ 2H^2\Omega\,\tilde{\rm D}_b\tilde{v}_a+ {1\over3H}\,\tilde{\rm D}_b\tilde{\rm D}_a \tilde{\vartheta}^{\prime}\nonumber\\ &&-{1\over3a^2H}\left(\tilde{\Delta}_{ab}^{\prime\prime} +\tilde{\mathcal{Z}}_{ab}^{\prime}\right)\,,  \label{tDbtva''}
\end{eqnarray}
where $\tilde{\Delta}_{ab}=a\tilde{\rm D}_b\tilde{\Delta}_a$ and $\tilde{\mathcal{Z}}_{ab}=a\tilde{\rm D}_b\tilde{\mathcal{Z}}_a$. The last two differential equations govern the linear kinematics of peculiar motions, as seen by observers ``living'' inside these bulk peculiar flows. Both relations apply to all scales and hold in an almost-FRW universe with nonzero background curvature and a pressureless (baryonic or/and CDM) matter. In what follows, we will attempt to extract analytical solutions from these formulae.

\subsection{Peculiar velocity}\label{ssPV}
To this point, we have considered a perturbed Friedmann universe without imposing any constraints on its spatial curvature. Hereafter, we will confine to an Einstein-de Sitter background by setting $\Omega=1$ and $H=2/3t$. However, Eqs.~(\ref{tv''2}) and (\ref{tDbtva''}) do not accept analytic solution even at the $\Omega=1$ limit. This is not uncommon in analytical cosmological studies and the standard way around it is to focus on the so-called long-wavelength solutions. In our case, the inhomogeneous components of (\ref{tv''2}) and (\ref{tDbtva''}) are comprised of spatial gradients in the peculiar volume expansion/contraction ($\tilde{\vartheta}$), in the matter density ($\tilde{\Delta}_a$) and in the universal expansion ($\tilde{\mathcal{Z}}_a$) -- of their first and second time-derivatives in particular. The effect of 3-gradients weakens and becomes less prominent as one moves to progressively larger scales. On these grounds, spatial gradients are typically dropped on scales close and beyond the Hubble horizon. This may reduce the range of the solutions, but provides important information regarding their large-scale behaviour and until horizon-crossing at least. Therefore, hereafter, our study will focus on the long-wavelength solutions.\footnote{As required, we applied the long-wavelength approximation to the final formulae and not at an earlier stage. This ensures that all the linear effects have been consistently incorporated into Eqs.~(\ref{tv''2}), (\ref{tDbtva''}).}

Isolating the homogeneous component of Eq.~(\ref{tv''2}), while setting $\Omega=1$ and $H=2/3t$ at the same time, we have
\begin{equation}
9t^2\tilde{v}_a^{\prime\prime}+ 3t\tilde{v}_a^{\prime}- 8\tilde{v}_a= 0\,,  \label{tv''3}
\end{equation}
on all scales where its inhomogeneous part is subdominant. The above accepts the power-law solution
\begin{equation}
\tilde{v}= \mathcal{C}_1t^{4/3}+ \mathcal{C}_2t^{-2/3}= \mathcal{C}_3a^2+ \mathcal{C}_4a^{-1}\,,  \label{tv1}
\end{equation}
which shows increase as $\tilde{v}\propto t^{4/3}\propto a^2$ (since $a\propto t^{2/3}$ after equipartition) for linear peculiar-velocity perturbations.\footnote{Evaluating the integration constants and using the cosmological redshift parameter ($z$) solution (\ref{tv1}) reads
\begin{eqnarray}
\hspace{-10pt} 3\tilde{v}= \left(\tilde{v}_0+{\tilde{v}_0^{\prime}\over H_0}\right) \left({1+z_0\over1+z}\right)^{\hspace{-2pt}2} \hspace{-2pt} +\left(2\tilde{v}_0-{\tilde{v}_0^{\prime}\over H_0}\right) \left({1+z\over1+z_0}\right)\,,  \label{tvz}
\end{eqnarray}
where the zero suffix marks the initial time. The above provides the peculiar velocity at redshift $z<z_0$, when one knows the initial conditions. These are typically set at decoupling, namely at $z_0=10^3$. According to (\ref{tvz}), essentially all the contribution to the residual peculiar velocity comes from higher redshifts, with $1<z<z_0$. Therefore, allowing for a late-time accelerated epoch (typically starting at $z<1$), should for all practical purposes leave our results unaffected.} Keeping only the growing mode, we deduce that $\tilde{v}/v_H\propto t^{5/3}\propto a^{5/2}$ after decoupling, with $v_H\propto a^{-1/2}$ giving the evolution of the Hubble velocity during the same period. These growth rates are significantly stronger than those reported in previous studies. Indeed, the Newtonian and quasi-Newtonian approaches give $v\propto t^{1/3}\propto a^{1/2}$ (e.g.~see~\cite{P,M} as well as~\cite{TF}), which implies $v/v_H\propto t^{2/3}\propto a$ (on an $\Omega=1$ background). On these grounds, the relativistic analysis presented here seems to favour bulk motions larger and faster than it is generally expected. An alternative interpretation of our results is that the observed peculiar velocities could have started considerably weaker than anticipated.

The analytic results given above apply to scales where the inhomogeneous component of (\ref{tv''2}) is sub-dominant. Hence, although one could readily apply solution (\ref{tv1}) beyond and near the Hubble radius, they should be cautious before doing so on scales well inside the horizon. There, the relativistic solution could reduce to its Newtonian counterpart, in which case it could also provide the initial conditions (near horizon-crossing) for a subsequent Newtonian treatment. Solving (\ref{tv''2}) on sub-Hubble lengths will probably require numerical treatment, with the analytical work providing the initial conditions. That aside, stronger growth-rates for the peculiar-velocity field on super-Hubble lengths, imply faster peculiar velocities at horizon-crossing. This in turn suggests higher residual values today and therefore provides theoretical support to a number of recent surveys reporting bulk peculiar flows larger and faster (sometimes considerably) than those typically expected~\cite{WFH}.

\subsection{Peculiar expansion/contraction}\label{ssPE/C}
Let us go back to Eq.~(\ref{tDbtva''}), which governs the linear evolution of the peculiar-velocity gradients. Assuming again a spatially flat FRW background, the trace of (\ref{tDbtva''}) gives
\begin{equation}
\tilde{\vartheta}^{\prime\prime}= -{5H\over2}\,\tilde{\vartheta}^{\prime}+ 2H^2\tilde{\vartheta}+ {1\over3H}\,\tilde{\rm D}^2\tilde{\vartheta}^{\prime}- {1\over3a^2H}\left(\tilde{\Delta}^{\prime\prime} +\tilde{\mathcal{Z}}^{\prime}\right)\,,  \label{tvtheta''}
\end{equation}
where $\tilde{\Delta}=\tilde{\Delta}_a{}^a$ and $\tilde{\mathcal{Z}}=\tilde{\mathcal{Z}}_a{}^a$ by default. Applying our long-wavelength approximation, we retain only the first two terms on the right-hand side of the above. Then, after equipartition, Eq.~(\ref{tvtheta''}) reads
\begin{equation}
9t^2\tilde{\vartheta}^{\prime\prime}+ 15t\tilde{\vartheta}^{\prime}- 8\tilde{\vartheta}= 0\,,  \label{thlstheta''}
\end{equation}
with a power-law solution of the form
\begin{equation}
\tilde{\vartheta}= \mathcal{C}_1t^{2/3}+ \mathcal{C}_2t^{-4/3}= \mathcal{C}_3a+ \mathcal{C}_4a^{-2}\,.  \label{SHtvtheta}
\end{equation}
Therefore, throughout the dust epoch, $\tilde{\vartheta}$ grows in tune with the dimensions of the universe, as long as the velocity perturbation remains outside the Hubble scale.

Including the Laplacian term does not prevent Eq.~(\ref{tvtheta''}) from accepting an analytic solution. Although, this may seem at odds with our adopted approximation scheme, it will allow us to probe (at least to a certain extent) the evolution of $\tilde{\vartheta}$ on sub-horizon scales as well. Then, applying a simple Fourier decomposition to the perturbation, leads to the differential equation
\begin{equation}
\tilde{\vartheta}_{(n)}^{\prime\prime}= -{5\over3t} \left[1+{2\over15}\left({\lambda_H\over\lambda_n}\right)^2\right] \tilde{\vartheta}_{(n)}^{\prime}+ {8\over9t^2}\,\tilde{\vartheta}_{(n)}\,,  \label{ththeta''}
\end{equation}
for the $n$-th harmonic mode.\footnote{We use the familiar harmonic splitting $\tilde{\vartheta}=\sum_n\tilde{\vartheta}_{(n)} \mathcal{Q}^{(n)}$, where $\tilde{\rm D}_a\tilde{\vartheta}_{(n)}=0$ and $\mathcal{Q}^{(n)}$ are scalar harmonic functions with $\dot{\mathcal{Q}}^{(n)}=0$ and $\tilde{\rm D}^2\mathcal{Q}^{(n)}=-(n/a)^2\mathcal{Q}^{(n)}$ (e.g.~see~\cite{TCM,EMM}).} Here, $\lambda_H=1/H$ is the Hubble radius, $\lambda_n=a/n$ is the physical scale of the perturbation (i.e.~of the bulk flow) and $n$ is the associated (comoving) wavenumber. The above has a scale-dependent solution. Indeed, introducing the scale parameter $\alpha=\lambda_H/\lambda_n$ -- with $\alpha>0$, we may express the solution of Eq.~(\ref{ththeta''}) in terms of the $\alpha$-parameter as
\begin{equation}
\tilde{\vartheta}= \mathcal{C}_1t^{\beta_1}+ \mathcal{C}_2t^{\beta_2}= \mathcal{C}_3a^{3\beta_1/2}+ \mathcal{C}_4a^{3\beta_2/2}\,,  \label{alphatvtheta}
\end{equation}
with
\begin{equation}
\beta_{1,2}=-{\alpha^2+3\mp\sqrt{a^4+6\alpha^2+81}\over9}\,.  \label{betas}
\end{equation}
Then, well outside the horizon, where $\alpha\ll1$, we recover solution (\ref{SHtvtheta}) with $\beta_1\simeq2/3$ and $\beta_2\simeq-4/3$. At horizon crossing, namely at the $\alpha=1$ threshold, one obtains $\beta_1=2(\sqrt{22}-2)/9$ and $\beta_2=-2(\sqrt{22}+2)/9$, which shows  a slight decrease in the growth-rate of $\tilde{\vartheta}$, relative to the super-Hubble solution. The slowing-down effect continuous as we move to progressively smaller lengths. In fact, on scales deep inside the Hubble radius (where $1/\alpha\rightarrow0$), a simple (linear) Taylor expansion of (\ref{betas}) gives $\beta_{1,2}\simeq -[\alpha^2+3\mp\alpha^2(1+3/\alpha^2)]/9$, with $\beta_1\simeq0$ and $\beta_2\simeq-2\alpha^2/9$. In other words, on sufficiently small scales the value $\tilde{\vartheta}$ tends to a constant.

The above results suggest that, after matter-radiation equality, the relative-strength ratio $\tilde{\vartheta}/H$ grows as $\tilde{\vartheta}/H\propto t^{5/3}\propto a^{5/2}$ on super-Hubble lengths. Well inside the horizon, on the other hand, we find that $\tilde{\vartheta}/H\propto t\propto a^{3/2}$. The divergence of the peculiar velocity field is typically related to the density contrast of the matter inhomogeneities (e.g.~see~\cite{P}). On these grounds, peculiar velocity perturbations that cross inside the horizon at decoupling, with $\tilde{\vartheta}/H\simeq10^{-5}$ at that time, could reach values as high as $\tilde{\vartheta}/H\sim10^{-1/2}$ today. Finally, we should note that our growth rates are again significantly stronger than the Newtonian ones, according to which $\vartheta/H\propto t^{2/3}\propto a$ (e.g.~see~\cite{P} as well as~\cite{TF}).

\subsection{Peculiar shear and vorticity}\label{ssPSV}
Taking the symmetric and traceless part of (\ref{tDbtva''}), we obtain the evolution formula of the peculiar shear. The skew component of the same relation, on the other hand, determines the propagation of the peculiar vorticity. More specifically, on an Einstein-de Sitter background, we have
\begin{equation}
\tilde{\varsigma}_{ab}^{\prime\prime}= -{5\over2}\,H\tilde{\varsigma}_{ab}^{\prime}+ 2H^2\tilde{\varsigma}_{ab}+ {1\over3H}\,\tilde{\rm D}_{\langle b}\tilde{\rm D}_{a\rangle}\tilde{\vartheta}^{\prime}- {1\over3a^2H}\left(\tilde{\Delta}_{\langle ab\rangle}^{\prime\prime} +\tilde{\mathcal{Z}}_{\langle ab\rangle}^{\prime}\right) \label{tvsigma''1}
\end{equation}
and
\begin{equation}
\tilde{\varpi}_{ab}^{\prime\prime}= -{5\over2}\,H\tilde{\varpi}_{ab}^{\prime}+ 2H^2\tilde{\varpi}_{ab}- {1\over3a^2H}\left(\tilde{\Delta}_{[ab]}^{\prime\prime} +\tilde{\mathcal{Z}}_{[ab]}^{\prime}\right)\,, \label{tvpi''}
\end{equation}
respectively.\footnote{The last term on the right-hand side of Eq.~(\ref{tvpi''}) stems from Frobenius' theorem (e.g.~see~\cite{W}), which ensures that rotating spacetimes do not possess integrable 3-dimensional hypersurfaces. Alternatively, one could say that the 4-velocity field is not hypersurface orthogonal in the presence of rotation. As a result, the (covariant) spatial gradients of scalars do not commute in rotating spaces, which in turn guarantees that $\tilde{\Delta}_{[ab]}$, $\tilde{\mathcal{Z}}_{[ab]}\neq0$. The interested reader is referred to~\cite{EBH} for further discussion and for an application of the Frobenius theorem to the relativistic study of cosmological perturbations.} When pressureless (baryonic or/and CDM) matter dominates, we have $a\propto t^{2/3}$ and $H=2/3t$. In such a case, the homogeneous component of Eq.~(\ref{tvsigma''1}) recasts as
\begin{equation}
9t^2\tilde{\varsigma}_{ab}^{\prime\prime}+ 15t\tilde{\varsigma}_{ab}^{\prime}- 8\tilde{\varsigma}_{ab}= 0\,. \label{tvsigma''2}
\end{equation}
The above accepts the power-law solution
\begin{equation}
\tilde{\varsigma}= \mathcal{C}_1t^{2/3}+ \mathcal{C}_2t^{-4/3}= \mathcal{C}_3a+ \mathcal{C}_4a^{-2}\,,  \label{tvsigma}
\end{equation}
which holds on sufficiently large scales where the inhomogeneous part of (\ref{tvsigma''1}) is subdominant (see also \S~\ref{ssPV} earlier). On these long wavelengths, the peculiar shear grows proportionally to the dimensions of the post-recombination host universe. The same is also true for the peculiar vorticity, since (after equipartition) the homogeneous component of (\ref{tvpi''}) accepts a power-law solution identical to (\ref{tvsigma''2}). Therefore, on scales where the higher-order derivatives in the last term of (\ref{tvpi''}) are negligible, the peculiar vorticity also grows as $\tilde{\varpi}\propto t^{2/3}\propto a$.\footnote{To the best of our knowledge, the only Newtonian treatment of the peculiar shear and vorticity was recently given in~\cite{TF}. There, it was found that $\tilde{\varsigma}/H \propto t^{2/3}\propto a$ and also that $\tilde{\varpi}/H\propto t^{-1/3}\propto a^{-1/2}$ after equipartition.}

These results imply that, after decoupling, the relative strength of the peculiar shear and that of the peculiar vorticity grows as $\tilde{\varsigma}/H,\,\tilde{\varpi}/H\propto t^{5/3}\propto a^{5/2}$. Therefore, on large enough scales (near and outside the Hubble horizon), peculiar flows could start with negligibly small amounts of shear and vorticity and still reach cosmologically relevant magnitudes today.

\section{Discussion}\label{sD}
Studies of cosmological peculiar motions have a research history that goes back several decades. Nevertheless, the great majority of the available theoretical work is essentially Newtonian in nature, despite the fact that the scales involved are a good fraction of the Hubble horizon. Moreover, all the aforementioned studies are performed in the CMB frame, although no real observer in the universe follows the smooth Hubble flow and despite the subtlety of the relative-motion effects. With these in mind, we have attempted a relativistic treatment of peculiar velocities in a ``tilted'' almost-FRW universe (with pressureless baryons or/and CDM) and conducted our analysis in the rest-frame of a typical galaxy (like our Milky Way), that moves relative to the universal expansion. Our main aim was to provide a better theoretical understanding of the bulk-motion kinematics. Further motivation came from an apparent disagreement between the bulk-flow surveys, with some reporting larger magnitudes and scales for the peculiar velocity fields than others (see~\cite{WFH} and~\cite{ND}, respectively, for representative though incomplete lists).

Even at the linear level, the kinematics and the dynamics of the universe appear different in the coordinate system of the bulk peculiar flow than in the frame of the Hubble expansion, simply because of relative motion effects. For instance, although the cosmic medium may look like a perfect fluid in the CMB frame, it will appear imperfect to the real observers solely due to their motion with respect to the smooth universal expansion~\cite{M}. Taking these fairly well known relativistic effects into account and employing linear (relativistic) cosmological perturbation theory, enabled us to obtain analytical expressions for the linear sources of peculiar velocities. This allowed us to go a step further than the quasi-Newtonian approach, where an effective gravitational potential and an evolution ansatz were introduced to address the issue.

The linear analysis confirmed that peculiar motions are the result of the ongoing structure formation process, and more specifically of the increasing inhomogeneity of the post-recombination universe. Our study also provided the first (to the best of our knowledge) relativistic insight to the evolution of the full peculiar kinematics. Technically speaking, this was achieved through a set of four differential equations, monitoring the linear propagation of the peculiar velocity itself, as well as those of the associated irreducible (local) kinematic quantities. The latter are the expansion/contraction, the shear and the rotation of the bulk flow. Solving the aforementioned differential formulae analytically, we found substantial growth for all aspects of the peculiar motion on sufficiently large scales. Moreover, the strength of the peculiar-velocity field, relative to the background Hubble expansion of the post-recombination universe, was found to increase in time as well. In particular, peculiar velocities were found to grow as $\tilde{v}\propto a^2$, a rate considerably stronger than the one reported in previous Newtonian studies (where $v\propto a^{1/2}$).  Stronger than the Newtonian rates were also obtained for the peculiar expansion/contraction, as well as for the peculiar shear and for the peculiar vorticity.

On theoretical grounds, the disagreement between the relativistic and the Newtonian results is due to the different way the two theories treat issues as fundamental as the gravitational field itself. General relativity, in particular, advocates that, in addition to the energy density and the pressure (isotropic and/or anisotropic), the energy flux gravitates as well. Applied to peculiar motions, this principle ensures a flux-contribution to the energy-momentum tensor that is entirely due to the (peculiar) motion of the matter. This then feeds into the conservation laws and eventually leads to expression (\ref{tAa2}), which (together with Eqs.~(\ref{tv''2}) and (\ref{tDbtva''})) plays a central role in this study and it can be seen as the relativistic correction to the Newtonian analysis (see \S~\ref{ssLRBTFs} and \S~\ref{ss4A} for details). The quasi-Newtonian approach and other relativistic studies have also bypassed the aforementioned input of the bulk-flow flux to the local gravitational field and the typical reasons are discussed in \S~\ref{ssPVs1} and \S~\ref{ss4A} respectively.

We finally remind the reader that we obtained our power-law solutions by confining to the long-wavelength limit of the associated differential equations. This means that the aforementioned analytic results apply on large enough scales, with the typical threshold set by the Hubble radius. On smaller scales, the solutions reported here should be treated with caution, though they can still provide the initial conditions for future (analytical or numerical) studies. At this stage, our relativistic analysis does suggest that the linear growth of large-scale peculiar velocities can be considerably stronger than it is generally expected. Indeed, even if these stronger growth-rates are confined to super-Hubble lengths, the residual peculiar-velocity field should be larger than anticipated. On these grounds, one should not be surprised to measure bulk flows larger and faster than it is generally expected. Like those reported in~\cite{WFH} for example.\\

\textbf{Acknowledgements:} ET was supported by a CNRS internship fellowship at the Institut d'Astrophysique de Paris (IAP), during the later stages of this work. CGT wishes to acknowledge support from a visiting fellowship by Clare Hall College and visitor support by DAMTP at the University of Cambridge, where part of this work was conducted. CGT also acknowledges support by the Hellenic Foundation for Research and Innovation (H.F.R.I.), under the “First Call for H.F.R.I. Research Projects to support Faculty members and Researchers and the procurement of high-cost research equipment grant” (Project Number: 789).


\begin{thebibliography}{}
\bibitem{Aetal} P.A.R. Ade, et al, Astron. Astrophys. \textbf{561}, A97 (2014).
\bibitem{P} P.J.E. Peebles, \textit{The Large-Scale Structure of the Universe} (Princeton University Press, Princeton, New Jersey, 1980); A. Nusser, A. Dekel, E. Bertschinger and G.R. Blumenthal, Astrophys. J. \textbf{379}, 6 (1991); T. Padmanabhan, \textit{Structure Formation in the Universe} (Cambridge University Press, Cambridge, 1993).
\bibitem{M} R. Maartens, Phys. Rev. D \textbf{58}, 124006 (1998); G.F.R. Ellis, H. van Elst and R. Maartens, Class. Quantum Grav. \textbf{18}, 5115 (2001).
\bibitem{TF} K. Filippou and C.G. Tsagas, [arXiv:2003.01186].
\bibitem{TCM} C.G. Tsagas, A. Challinor and R. Maartens, Phys. Rep. \textbf{465}, 61 (2008).
\bibitem{EMM} G.F.R. Ellis, R. Maartens and M.A.H. MacCallum, \textit{Relativistic Cosmology} (Cambridge University Press, Cambridge, 2012).
\bibitem{WFH} R. Watkins, H.A. Feldman and M.J. Hudson, Mon. Not. R. Astron. Soc. \textbf{392}, 743 (2009); A. Kashlinsky, F. Atrio-Barandela, D. Kocevski and H. Ebeling, Astrophys. J. \textbf{686}, L49 (2009); G. Lavaux, R.B. Tully, R. Mohayaee, S. Colombi, Astrophys. J. \textbf{709}, 483 (2010);  J. Colin, R. Mohayaee, S. Sarkar and A. Shafieloo, Mon. Not. R. Astron. Soc. \textbf{414}, 264 (2011); C. Magoulas et al, \textit{IAU Symposium} \textbf{308}, 336 (2016); A. Salehi, M. Yarahmadi and S. Fathi, [arXiv:2001.01743]; R. Mohayaee, M. Rameez and S. Sarkar [arXiv:2003.10420].
\bibitem{KE} A.R. King and G.F.R. Ellis, Commun. Math. Phys. \textbf{31}, 209, (1973).
\bibitem{Eh} J.  Ehlers, Akad. Wiss. Lit. Mainz Abh. Math.-Nat. Kl. \textbf{11}, 792 (1961).
\bibitem{ET} G.F.R. Ellis and C.G. Tsagas, Phys. Rev. D \textbf{66}, 124015 (2002); C.G. Tsagas and M.I. Kadiltzoglou, Phys. Rev. D \textbf{88}, 083501 (2013).
\bibitem{SW} J.M. Stewart and M. Walker, Proc. R. Soc. A \textbf{341}, 49 (1974).
\bibitem{TK} C.G. Tsagas and M.I. Kadiltzoglou, Phys. Rev. D \textbf{92}, 043515 (2015).
\bibitem{E} G.F.R. Ellis, Mon. Not. R. Astron. Soc. \textbf{243}, 509 (1990).
\bibitem{CMV-R} J.A.R. Cembranos, A.L. Maroto and H. Villarrubia-Rojo, JCAP \textbf{06}, 041 (2019).
\bibitem{RV} G. Rigopoulos and W. Valkenburg, Phys. Rev. D \textbf{86}, 043523 (2012).
\bibitem{MBBM} I. Milillo, Bertacca D., M. Bruni and A. Maselli, Phys. Rev. D \textbf{92}, 023519 (2015).
\bibitem{W} R.M. Wald, \textit{General Relativity} (Chicago University Press, Chicago, 1984); E. Poisson, \textit{A Relativist's Toolkit} (Cambridge University Press, Cambridge, 2004).
\bibitem{EBH} G.F.R. Ellis, M. Bruni and J. Hwang, Phys. Rev. D \textbf{42}, 1035 (1990).
\bibitem{ND} A. Nusser and M. Davis, Astrophys. J. \textbf{736}, 93 (2011); S.J. Osborne, D.S.Y. Mak, S.E. Church and E, Pierpaoli, Astrophys. J. \textbf{737}, 98 (2011); E. Branchini, M. Davis and A. Nusser, Mon. Not. R. Astron. Soc. \textbf{424}, 472 (2012); S.J. Turnbull et al, Mon. Not. R. Astron. Soc. \textbf{420}, 447 (2012); Ma Y.-Z. and Pan J., Mon. Not. R. Astron. Soc. \textbf{437}, 1996 (2014).
\end{thebibliography}
\end{document}